\documentclass[12pt]{article}
\usepackage{epsfig}
\usepackage{amsfonts}

\def\beq{\begin{equation}}
\def\eeq#1{\label{#1}\end{equation}}
\def\eeqn{\end{equation}}

\def\beqa{\begin{eqnarray}}
\def\eeqa#1{\label{#1}\end{eqnarray}}
\def\eeqan{\end{eqnarray}}

\let\bar=\overbar

\def\tr{{\mbox{\rm tr}}}

\def\Dslash{\not{\hbox{\kern-4pt $D$}}}
\def\dslash{\not{\hbox{\kern-2pt $\del$}}}

\def\msb{{\bar{\ssstyle M \kern -1pt S}}}

\newcommand{\dde}[1]{\partial_{#1}}
\newcommand{\la}{\langle}
\newcommand{\ra}{\rangle}

\newcommand{\diag}{{\rm diag}}

\newcommand{\GeV}{{\rm GeV}}

\renewcommand{\Re}{{\rm Re\,}}

\def\Title#1{\begin{center} {\Large {\bf #1} } \end{center}}

\begin{document}

\Title{Euclidean--Minkowskian Duality of Wilson--Loop Correlation Functions}

\bigskip\bigskip

\begin{raggedright}

{\it Matteo Giordano{\footnote{Speaker at the conference.}} and Enrico Meggiolaro
\index{Giordano, M.}\\ 
Dipartimento di Fisica -- Universit\`a di Pisa, \\
and INFN -- Sezione di Pisa, \\
Largo Pontecorvo 3,
I-56127 Pisa, Italy}
\bigskip\bigskip
\end{raggedright}

\begin{abstract}
We discuss the analyticity properties of the Wilson--loop correlation
    functions relevant to the problem of {\it soft} high--energy scattering,
    directly at the level of the functional integral, in a genuinely
    nonperturbative way.
\end{abstract}

\section{Introduction}

Since the seminal paper~\cite{Nachtmann91} by O.~Nachtmann, much work
has been done on the problem of {\it soft} high--energy scattering in
strong interactions in the framework of nonperturbative QCD (for a review
see Ref.~\cite{Dosch}). 
In this approach, at high center--of--mass energy and small transferred
momentum ($\sqrt{|t|} \le 1\, \GeV \ll \sqrt{s} $) hadronic scattering
amplitudes are reconstructed from the elastic scattering amplitude of
the corresponding system of constituent partons, after folding with the
appropriate hadron wave functions. In particular, in the case of
meson--meson elastic scattering, the corresponding amplitudes can be
reconstructed  from the
correlation functions of two Wilson loops (which describe the
scattering of two colour dipoles of fixed transverse size) running along
the trajectories of the colliding
hadrons~\cite{DFK} (see 
also~\cite{Dosch}). As an infrared (IR)
regularisation~\cite{Verlinde}, the loops are taken to form
a finite hyperbolic angle $\chi$ in Minkowski spacetime, and moreover
they are taken to be of finite length $2T$; the physical amplitudes
(which are expected to be IR finite~\cite{BL}) are recovered in the
limit $T\to\infty$ at large $\chi{\simeq}\log(s/m^2)$ (for
${s\to\infty}$).  

It has been shown
in~\cite{Meggiolaro97,Meggiolaro98,Meggiolaro02,Meggiolaro05,crossing,Meggiolaro07} 
that, under certain analyticity hypotheses, the relevant correlation
functions can be reconstructed from the ``corresponding'' correlation
functions of two Euclidean Wilson loops, of finite
length $2T$, and forming an angle $\theta$ in Euclidean space, by
means of the double analytic continuation $\theta \to -i\chi$, $T\to
iT$. This {\it Euclidean--Minkowskian duality} of Wilson--loop
correlation functions has made possible to approach the problem of
{\it soft} high--energy scattering with the nonperturbative techniques
of Euclidean Quantum Field Theory, 
such as the {\it Instanton Liquid Model}~\cite{instanton1}, the
{\it Stochastic Vacuum Model}~\cite{LLCM2}, the {\it AdS/CFT
correspondence}~\cite{JP}, and {\it Lattice Gauge 
Theory}~\cite{lattice}. However, until recent times,
the {\it analytic--continuation relations}, although expected to be an
exact result, had only been verified in perturbation
theory~\cite{Meggiolaro97,Meggiolaro05,crossing,BB}, while a
nonperturbative foundation was lacking (except in the case of {\it
quenched} QED, where an exact calculation can be performed both in 
the Euclidean and Minkowskian theories~\cite{Meggiolaro05}).

In Ref.~\cite{EMduality} we have argued, on nonperturbative
grounds, that the required analyticity hypotheses 
are indeed satisfied. The strategy we have used is to push the dependence on
the relevant variables into the action by means of a field and coordinate 
transformation, and then to allow them to take complex values.  
In particular, we have determined the analyticity domain of the
relevant Euclidean correlation function, and we have shown that the
corresponding Minkowskian quantity is recovered with the usual double
analytic continuation in $\theta$ and $T$ inside this domain;
moreover, the extra conditions that allow one to derive the {\it
  crossing--symmetry relations} found in Ref.~\cite{crossing} have
been shown to be satisfied, and we have refined the argument given
in Ref.~\cite{Meggiolaro05} for the analytic continuation of the
correlation function with the IR cutoff removed. The formal
manipulations of the functional integral used to obtain these results
have been justified making use of a lattice regularisation.

\section{High--energy meson--meson scattering and\\ Wilson--loop
  correlation functions} 

The elastic scattering amplitudes of two mesons (taken for simplicity
with the same mass $m$) in the {\it soft} high--energy regime can be
reconstructed, after folding with the appropriate wave functions, from
the scattering amplitude ${\cal M}_{(dd)}$ of two dipoles of fixed
transverse size $\vec{R}_{i\perp}$, and fixed longitudinal momentum
$f_i$ of the two quarks in the two dipoles, respectively
($i=1,2$)~\cite{DFK}:
\beq
  {\cal M}_{(dd)} (s,t;1,2) 
\equiv -i~2s \displaystyle\int d^2 \vec{z}_\perp
e^{i \vec{q}_\perp \cdot \vec{z}_\perp}
{\cal C}_M(\chi;\vec{z}_\perp;1,2) ,
\eeq{scatt-loop}
where the arguments ``$i$'' stand for ``$\vec{R}_{i\perp},
f_i$'',
$t =
-|\vec{q}_\perp|^2$ ($\vec{q}_\perp$ being the transferred momentum)
and $s=2m^2(1+\cosh\chi)$.
The correlation function ${\cal C}_M$ is defined as the
limit $\displaystyle {\cal C}_M \equiv \lim_{T\to\infty} {\cal G}_M $
of the correlation function of two loops of finite length $2T$,
\beq
  {\cal G}_M(\chi;T;\vec{z}_\perp;1,2) \equiv
{ \langle {\cal W}^{(T)}_1 {\cal W}^{(T)}_2 \rangle \over
\langle {\cal W}^{(T)}_1 \rangle
\langle {\cal W}^{(T)}_2 \rangle } - 1,
\eeq{GM}
where $\langle\ldots\rangle$ are averages in the sense of the QCD
functional integral, and
\beq
{\cal W}^{(T)}_{1,2} \equiv
{\frac{1}{N_c}} \tr \left\{ {\cal P} \exp
\left[ -ig \displaystyle\oint_{{\cal C}_{1,2}} A_{\mu}(x) dx^{\mu} \right]
\right\} 
\eeq{QCDloops}
are Wilson loops in the fundamental representation of $SU(N_c)$; the
paths are made up of the classical trajectories of quarks and
antiquarks, 
\beq
{\cal C}_1 :
X^{1q[\bar{q}]}_{}(\tau)
 = z + {p_{1} \over m} \tau + 
f^{q[\bar{q}]}_1 R_{1} , \quad
{\cal C}_2 :
X^{2q[\bar{q}]}_{}(\tau)
 = {p_{2} \over m} \tau + 
f^{q[\bar{q}]}_2 R_{2},
\eeq{traj}
with $\tau\in [-T,T]$, and closed by straight--line paths in the
transverse plane at $\tau=\pm T$ in order to ensure gauge invariance. 
Here
\beq 
{p_1}={m}\Big( \cosh \displaystyle\frac{\chi}{2},\sinh \frac{\chi}{2},\vec{0}_\perp \Big) ,~~~
{p_2}={m}\Big( \cosh \displaystyle \frac{\chi}{2},-\sinh
\frac{\chi}{2},\vec{0}_\perp \Big), 
\eeq{p1p2}
and moreover, $R_1 = (0,0,\vec{R}_{1\perp})$, $R_2 = (0,0,\vec{R}_{2\perp})$,
$z = (0,0,\vec{z}_\perp)$, and $f^{q}_i = 1-f_i$, $f^{\bar{q}}_i = -f_i$
($i=1,2$), with $f_i$ the longitudinal momentum fraction of quark $i$,
$f_i\in [0,1]$. 
The Euclidean counterpart of Eq.~(\ref{GM}) is
\beq
  {\cal G}_E(\theta;T;\vec{z}_\perp;1,2) \equiv
\frac{ \langle \widetilde{\cal W}^{(T)}_1 \widetilde{\cal W}^{(T)}_2 \rangle_E}
{\langle \widetilde{\cal W}^{(T)}_1 \rangle_E
\langle \widetilde{\cal W}^{(T)}_2 \rangle_E } - 1,
\eeq{GE}
where now $\langle\ldots\rangle_E$ is the average in the sense of the
Euclidean QCD functional integral, and the Euclidean Wilson loops
\beq
\widetilde{\cal W}^{(T)}_{1,2} \equiv
{\displaystyle\frac{1}{N_c}} \tr \left\{ {\cal P} \exp
\left[ -ig \displaystyle\oint_{\widetilde{\cal C}_{1,2}} A_{E\mu}(x_E) dx_{E\mu} \right]
\right\} 
\eeq{QCDloopsE}
are calculated on the following straight--line paths,
\beq
\widetilde{\cal C}_1 :
X^{1q[\bar{q}]}_{E}(\tau)
 = z + {p_{1E} \over m} \tau + 
f^{q[\bar{q}]}_1 R_{1E} , \quad
\widetilde{\cal C}_2 :
X^{2q[\bar{q}]}_{E}(\tau)
 = {p_{2E} \over m} \tau + 
f^{q[\bar{q}]}_2 R_{2E},
\eeq{trajE}
with $\tau\in [-T,T]$, and closed by straight--line paths in the
transverse plane at $\tau=\pm T$. Here\footnote{For convenience, we
  take the Euclidean indices to run from 0 to 3, and we take the
  ``Euclidean time'' to be the zero--th Euclidean coordinate.}
\beqa
{p_{1E}}={m}\Big(\cos\frac{\theta}{2},\sin\frac{\theta}{2},\vec{0}_{\perp}\Big), 
\quad
{p_{2E}}={m}\Big(\cos\frac{\theta}{2},-\sin\frac{\theta}{2},\vec{0}_{\perp}\Big), 
\eeqa{p1p2E}
and $R_{iE} = (0,0,\vec{R}_{i\perp})$, $z_E = (0,0,\vec{z}_{\perp})$
(the
transverse vectors are taken to be equal in the two
cases).
Again, we define the correlation function with the IR cutoff removed
as $\displaystyle {\cal C}_E \equiv \lim_{T\to\infty} {\cal G}_E $. 

It has been shown
in~\cite{Meggiolaro97,Meggiolaro98,Meggiolaro02,Meggiolaro05} that 
the correlation functions in the two theories are connected by the
{\it analytic--continuation relations}
\beqa
{\cal G}_M(\chi;T;\vec{z}_\perp;1,2)
&=& \overline{\cal G}_E (-i\chi;iT;\vec{z}_\perp;1,2) ,
\qquad \forall\chi\in {\cal I}_{M},\nonumber \\
  {\cal G}_E(\theta;T;\vec{z}_\perp;1,2)
&=& \overline{\cal G}_M (i\theta;-iT;\vec{z}_\perp;1,2) ,
\qquad \forall\theta\in {\cal I}_{E}.
\eeqa{analytic}
Here we denote with an overbar the analytic extensions of the
Euclidean and Min\-kow\-skian correlation functions, starting from the
real intervals  ${\cal I}_E\equiv (0,\pi)$ and
${\cal I}_M \equiv (0,\infty)$ of the respective angular variables,
with positive real $T$ in both cases, into domains of the complex
variables $\theta$ 
(resp.~$\chi$) and $T$ in a two--dimensional complex space; these
domains are assumed to contain the intervals  $-i{\cal I}_M$ (at
positive imaginary $T$)\footnote{We use here and in the following the
  notation $\alpha +   \beta{\cal I} =   \{ \alpha + \beta z | z\in
  {\cal I} \}$.} and  $i{\cal I}_E$ (at negative imaginary $T$) in 
the two cases, respectively. Due to the symmetries of the two
theories, the restriction to ${\cal I}_M$ and ${\cal I}_E$ does not
imply any restriction on the physical content of the correlation
functions (see Ref.~\cite{crossing}).
Under certain analyticity hypotheses in the $T$ variable, the
following relations are obtained for the correlation functions with
the IR cutoff $T$ removed~\cite{Meggiolaro05}:
\beqa
{\cal C}_M(\chi;\vec{z}_\perp;1,2) &=&
\overline{\cal C}_E(-i\chi;\vec{z}_\perp;1,2) ,
\qquad \forall\chi\in {\cal I}_M,\nonumber \\
  {\cal C}_E(\theta;\vec{z}_\perp;1,2) &=&
\overline{\cal C}_M(i\theta;\vec{z}_\perp;1,2) ,
\qquad \phantom{-}\forall\theta\in {\cal I}_E.
\eeqa{analytic_C}
Finally, we recall the {\it crossing--symmetry
  relations}~\cite{crossing} 
\beqa
&\overline{\mathcal{G}}_M(i\pi-\chi;T;\vec{z}_{\perp};1,2)
=\mathcal{G}_M(\chi;T;\vec{z}_{\perp};1,\overline{2}) 
=\mathcal{G}_M(\chi;T;\vec{z}_{\perp};\overline{1},2) ,
\quad \forall\chi\in {\cal I}_M,&
\nonumber \\
&  \mathcal{G}_E(\pi-\theta;T;\vec{z}_{\perp};1,2)
=\mathcal{G}_E(\theta;T;\vec{z}_{\perp};1,\overline{2}) 
~=\mathcal{G}_E(\theta;T;\vec{z}_{\perp};\overline{1},2) ,
\quad\forall\theta\in {\cal I}_E ,&
\eeqa{eq:crossrel}
that hold for every positive real $T$, and thus also for the
correlation functions 
with the IR cutoff removed; 
here the arguments ``$\overline{i}$'' 
stand for ``$-\vec{R}_{i\perp}, 1-f_i$''.
The Euclidean relation in
(\ref{eq:crossrel}) holds without any analyticity hypothesis, while in
the Min\-kow\-skian case the analyticity domain for the analytic
extension $\overline{\cal G}_M$ should include also the interval (in the
complex--$\chi$ plane) ${\cal  I}_M^{(c)}= i\pi - {\cal I}_M$ (for
positive real $T$), where the physical amplitude for the  ``crossed''
channel is then expected to lie. 

\section{Field and coordinate transformation}

To address the issue of the analytic extension of the correlation
functions to complex values of the angular variables and of $T$, we
shall appropriately rescale the coordinates and fields, in order for
the dependence on the relevant variables to drop from the Wilson--loop
operators, and to move into the action. 
We consider here the pure--gauge case only; the inclusion of fermions
(at the formal level) is discussed in~\cite{EMduality}.

We first rescale~\cite{Meggiolaro98,Meggiolaro02,Meggiolaro05}
$\tau\to \alpha\tau$ in the ${\cal P}$--exponentials 
corresponding to the longitudinal sides, setting $\alpha= T/T_0$ with
$T_0$ some fixed time (length) scale: in this way one shows explicitly
that the loops depend on $T$ only through the
combinations $(T/T_0) p_i/m$ and $(T/T_0)
p_{Ei}/m$. Next, we rescale coordinates and fields as follows. To
unify the treatment of the Euclidean and Minkowskian cases we use the
same symbol $\phi_\mu$ for the transformed gauge fields, and $y^\mu$ for
the transformed coordinates (we can use upper indices for the new
coordinates also in the Euclidean case without ambiguity). We then set
\beqa
  y^{\mu} = M^{\mu}_{\phantom{\mu}\nu}x^{\nu},&  \qquad &
  A_{\mu}(x) = \phi_{\nu}(y)M^{\nu}_{\phantom{\nu}\mu},\nonumber\\
  y^{\mu} = M_{E\mu\nu}x_{E\nu}, & \qquad &
  A_{E\mu}(x_E) = \phi_{\nu}(y)M_{E\nu\mu},
\eeqa{eq:ME-transf}
in the Minkowskian and Euclidean cases, respectively, 
where $M$ and $M_E$ are the diagonal matrices
\beqa
  M^{\mu}_{\phantom{\mu}\nu} &=&
  \diag\left(\frac{T_0}{T}\frac{1}{\sqrt{2}\cosh(\chi/2)},
  \frac{T_0}{T}\frac{1}{\sqrt{2}\sinh(\chi/2)},1,1\right),\nonumber\\
  M_{E\mu\nu} &=&
  \diag\left(\frac{T_0}{T}\frac{1}{\sqrt{2}\cos(\theta/2)},
  \frac{T_0}{T}\frac{1}{\sqrt{2}\sin(\theta/2)},1,1\right).
\eeqa{eq:matrix}
The Wilson loops in the two theories are then changed into the same
functionals of the new variables; however, the transformed actions are
different in the two cases, and so are the expectation
values. To make this clear, we introduce the notation
\begin{equation}
  \label{eq:fintegral}
  \la {\cal O}[\phi] \ra_S \equiv Z_S^{-1}{\displaystyle\int[D\phi]{\cal
  O}[\phi] e^{-S[\phi]}}, \quad Z_S={\displaystyle\int[D\phi]e^{-S[\phi]}},
\end{equation}
and we denote with
\beq
W_{\Gamma_{1,2}}[\phi] \equiv  {\frac{1}{N_c}} \tr \left\{ {\cal P} \exp
\left[ -ig \displaystyle\oint_{{\Gamma}_{1,2}} \phi_{\mu}(y) dy^{\mu} \right]\right\}
\eeq{}
the new Wilson loops. 
Here the new paths are given by
\beqa
{\Gamma}_1 &:& \quad
Y_{1q[\bar{q}]}^\mu(\tau) = z^\mu +
\frac{\delta^{\mu}_{\phantom{\mu}0}+\delta^{\mu}_{\phantom{\mu}1}}{\sqrt{2}}
\tau + f_1^{q[\bar{q}]} R_1^\mu, ~~~~  \nonumber \\
{\Gamma}_2 &:& \quad
Y_{2q[\bar{q}]}^\mu(\tau) =
\frac{\delta^{\mu}_{\phantom{\mu}0}-\delta^{\mu}_{\phantom{\mu}1}}{\sqrt{2}}
\tau + f_2^{q[\bar{q}]} R_2^\mu,
\eeqa{}
with $\tau\in [-T_0,T_0]$, and closed by the usual
transverse straight--line paths at $\tau= \pm T_0$. 
We then write for the correlation functions and expectation values in
the two theories
\beqa
  \la {\cal W}^{(T)}_{1}{\cal W}^{(T)}_{2} \ra &=& \la W_{\Gamma_1}
  W_{\Gamma_2} \ra_{-iS_M^{\rm Y.M.}}, \qquad
  \la {\cal W}^{(T)}_{i} \ra = \la W_{\Gamma_i} \ra_{-iS_M^{\rm Y.M.}}, \nonumber\\ 
  \la \widetilde{\cal W}^{(T)}_{1}\widetilde{\cal W}^{(T)}_{2} \ra_E &=&
  \la W_{\Gamma_1} W_{\Gamma_2} \ra_{S_E^{\rm Y.M.}}, \qquad
  \la \widetilde{\cal W}^{(T)}_{i} \ra_E = \la W_{\Gamma_i} \ra_{S_E^{\rm Y.M.}} ,
\eeqa{eq:trans_corrfunc}
where $S_M^{\rm Y.M.}$ and $S_E^{\rm Y.M.}$ are the transformed Minkowskian and Euclidean
pure--gauge (Yang--Mills) actions:
\beqa
  \label{eq:mod_M_ac}
  S_M^{\rm Y.M.} &=& -\sum_{\mu,\nu=0}^3  C_{M\mu\nu}(\chi,T)\frac{1}{2}\int d^4y\,
  \tr(\Phi_{\mu\nu})^2\\ 
  \label{eq:mod_E_ac}
  S_E^{\rm Y.M.} &=& \phantom{-}\sum_{\mu,\nu=0}^3 {C}_{E\mu\nu}(\theta,T)\frac{1}{2}\int d^4y\,
  \tr(\Phi_{\mu\nu})^2.
\eeqa{}
Here $(\Phi_{\mu\nu})^2$ is understood as the square of the Hermitian
matrix 
 $ \Phi_{\mu\nu} = \dde{\mu}\phi_{\nu} - \dde{\nu}\phi_{\mu} +
  ig[\phi_{\mu},\phi_{\nu}]$, 
and the symmetric coefficients $C_{M\mu\nu}$ and ${C}_{E\mu\nu}$
are ($\nu_\perp=2,3$)
\beqa
&\displaystyle  C_{M01}  = -C_{M23}^{-1} = -\left(\frac{T_0}{T}\right)^2\frac{1}{|\sinh\chi|},\quad
  C_{M0\nu_{\perp}}  =  -C_{M1\nu_{\perp}}^{-1} = -\frac{|\sinh\chi|}{\cosh\chi+1}&
\label{coeff_1}\\
&\displaystyle  {C}_{E01}  = {C}_{E23}^{-1}  =
  \left(\frac{T_0}{T}\right)^2\frac{1}{|\sin\theta|}, \quad
  {C}_{E0\nu_{\perp}}  = {C}_{E1\nu_{\perp}}^{-1} =  \frac{|\sin\theta|}{\cos\theta+1},&
\label{coeff_2}
\eeqa{}
and $C_{M\mu\mu}={C}_{E\mu\mu}=0$ $\forall \mu$. 

Restricting the
angular variables to the intervals $\chi\in {\cal I}_M$ 
and $\theta\in {\cal I}_E$ 
(see remark above), 
we can drop the absolute  values in Eqs.~(\ref{coeff_1}) and
(\ref{coeff_2}), obtaining 
coefficient functions which can be analytically  
extended throughout the respective complex planes in both
variables, with the possible exception (depending on the specific
coefficient) of the isolated singular points (poles)
$T=0,\infty$, and $\chi=0,\infty$ in the Minkowskian case or
$\theta=n\pi$, $n\in\mathbb{Z}$ in the Euclidean case.  To avoid
confusion, 
we will denote with an overbar the {\it analytic extensions}
$\overline{C}_{M\mu\nu}$ and $\overline{C}_{E\mu\nu}$ obtained
starting from ${\cal I}_M$ and ${\cal I}_E$ at real positive $T$, in
the two cases respectively.

\section{Analyticity domain of the Euclidean correlation function}

A function of a complex variable is analytic if its derivative exists
in complex sense. If we were allowed to bring the derivative under the sign
of integral, we could infer the analytic properties of the correlation
function ${\cal G}_E$ directly from its functional--integral
representation. 
Here we will give a formal argument, based on the
following assumption: one can bring the derivative with respect to a
parameter under the functional integral sign as long as the resulting
integral is convergent, in analogy with the case of ordinary
integrals\footnote{Strictly speaking, in the latter case a sufficient
  condition is {\it uniform} convergence.}. 

The functional integrals defined by means of
Eq.~(\ref{eq:fintegral}) are expected to be convergent as long as
the real part of the action is positive--definite; 
allowing for derivatives to pass under the sign of
integral, it is easy to see that the convergence properties of the
functional integral are left unchanged, and so 
we conclude (formally) that
the correlation function ${\cal G}_E$ can be analytically extended to
complex values of $\theta$ and $T$ for which the real part of the
action $S_E^{\rm Y.M.}$ is positive--definite, and this happens if and
only if the {\it convergence conditions}  
\begin{equation}
  \label{eq:convcond}
  \Re\overline{C}_{E\mu\nu}(\theta,T) > 0 \quad \forall \mu,\nu
\end{equation}
for the (analytically extended) coefficients are satisfied.
Note that singular points of the coefficients are artifacts of the
field and coordinate transformation, and they are not necessarily
singular points of the correlation function. 
Indeed, while singularities are
expected at the points $\theta=0,\pi$ on the basis of the
relation between the correlation function ${\cal G}_E$ and the static
dipole--dipole potential~\cite{Pot} (see also~\cite{crossing}), no
singularity is expected at $T=0$, where ${\cal G}_E$ is expected to
vanish. 

Substituting $\theta$ with the complex variable $z\equiv\theta-i\chi$
(with real $\theta$ and $\chi$) and writing for the complex variable
$T$, $T=|T|e^{i\psi/2}$, one easily sees that the {\it convergence
  conditions} (\ref{eq:convcond}) are equivalent to
\beqa
  && F(\theta,\chi,\psi) \equiv e^{\chi}\sin(\theta+\psi) +
  e^{-\chi}\sin(\theta-\psi) > 0, \nonumber\\
  && \sin\theta(\cosh\chi + \cos\theta) > 0.
\eeqa{} 
It has been shown in~\cite{EMduality} that the previous inequalities 
define a connected subset ${\cal V}$ of the
3D real $(\theta,\chi,\psi)$--space; moreover, as the modulus
$|T|$ never enters the previous equations, the section of the
analyticity domain is the same irrespectively of $|T|$. No dependence on
the arbitrary parameter $T_0$ is found, too, as expected. One then
finds a connected analyticity domain ${\cal D}_E$,
\begin{equation}
  \label{eq:domain}
  {\cal D}_E  = \{(z,T)\in \mathbb{C}^2 | (\theta,\chi,\psi)\in {\cal V}\}
\end{equation}
for the extension of the Euclidean correlation function from $\theta\in{\cal
  I}_E$ at positive real $T$. 
Note that the analyticity domain must be symmetric under $z\to z^*$ and under
$z\to \pi-z$, as one can easily show taking into account the identities
\beq
F(\theta,\chi,\psi) =  F(\theta,-\chi,-\psi) =
F(\pi-\theta,-\chi,\psi).
\eeq{}
\begin{figure}[tb]
\begin{center}
\epsfig{file=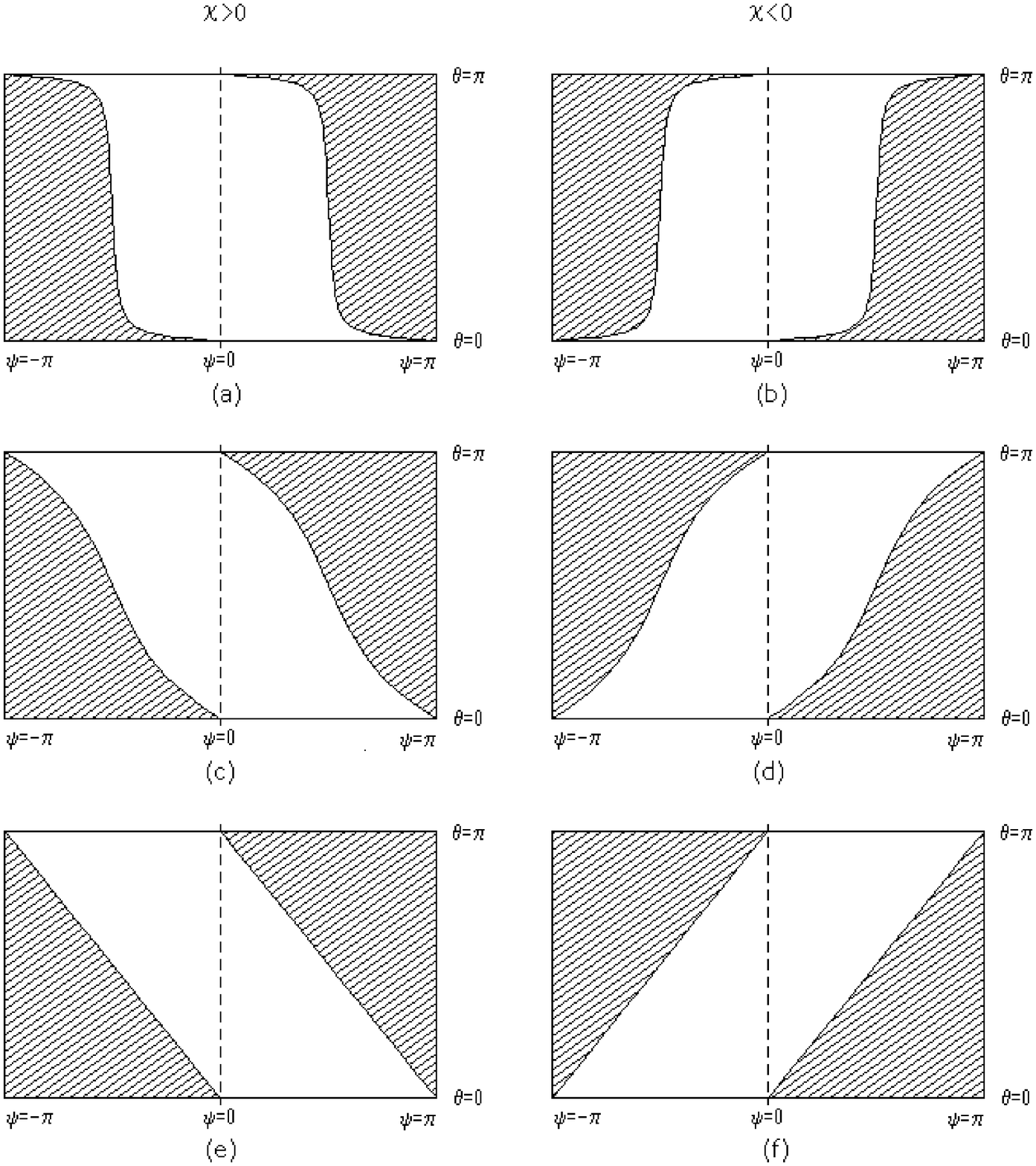,height=3.in} \raisebox{0.7in}{\hspace{0.6cm}\epsfig{file=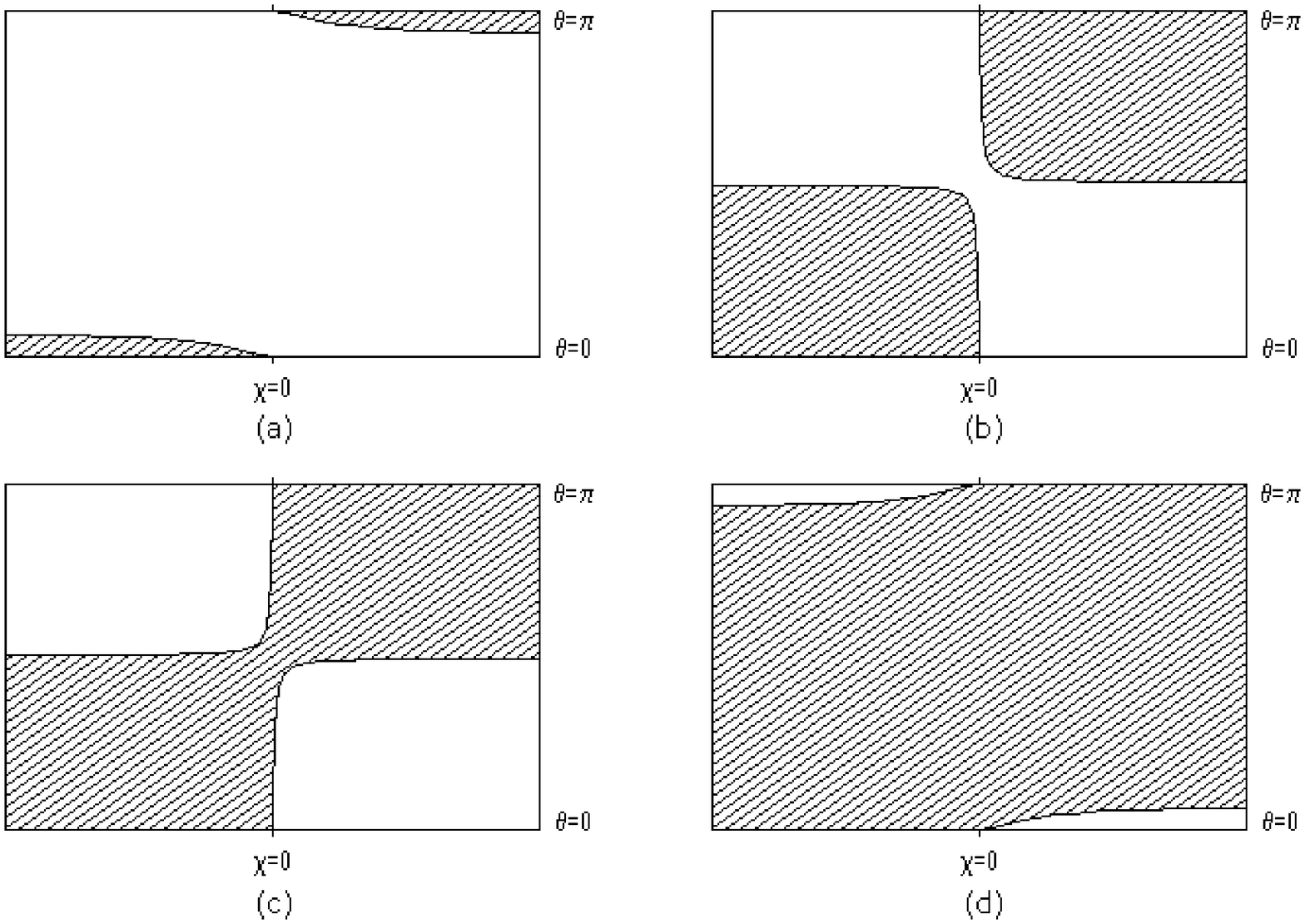,height=1.7in}} 
\caption{Sections of the analyticity domain (white area) at fixed $\chi$ (left) and at
  fixed $\psi$ (right).}
\label{fig:fixchi}
\end{center}
\end{figure}
Sections of this subset at fixed $\chi$ are shown in
Fig.~\ref{fig:fixchi} (left). Note that the domain ``thins out''
as one tends toward the ``physical''
edges $E^{\rm dir}$ and $E^{\rm
  cross}$, 
\beq
E^{\rm dir/cross} =  \{(z,T)\in \mathbb{C}^2\, |\,
\theta=0/\pi,\chi\in\mathbb{R}^{+/-},\psi=\pi\},
\eeq{}
and also towards the other two edges $E^{\rm
  dir}{}^*$ and $E^{\rm cross}{}^*$ [here $E^* = \{(z,T) |
(z^*,T^*)\in E\}$].
Sections of the same analyticity domain at fixed $\psi$ are shown in
Fig.~\ref{fig:fixchi} (right): the whole ``strip'' ${\cal 
  S}_E\equiv\{z=\theta-i\chi\, |\, \theta\in
(0,\pi),\chi\in\mathbb{R}\}$ (at $\psi=0$) reduces to disjoint regions 
near the edges of the domain (at $\psi\simeq\pm\pi$). 

As we approach $E^{\rm dir}$ from the inside, 
the coefficients $\overline{C}_{E\mu\nu}$ become imaginary,
and 
\begin{equation}
  \overline{C}_{E\mu\nu}(-i\chi,iT) =
  i{C}_{M\mu\nu}(\chi,T) 
\end{equation}
so that $S_E^{\rm Y.M.} \stackrel{\theta\to -i\chi,\,T\to
  iT}{\longrightarrow} -iS_M^{\rm Y.M.}$, 
i.e., according to Eq.~(\ref{eq:trans_corrfunc}),
\begin{equation}
  {\cal G}_M (\chi;T;\vec{z}_\perp;1,2)=
  \overline{\cal G}_E(-i\chi;iT;\vec{z}_\perp;1,2), 
  \qquad \forall \chi\in{\cal I}_M,T\in\mathbb{R}^+.
\label{eq:an_cont_phys}
\end{equation}
We thus find that Minkowskian and Euclidean correlation functions are
connected by the expected analytic
continuation~\cite{Meggiolaro97,Meggiolaro98,Meggiolaro02,Meggiolaro05},
of which we have given here an alternative derivation. 
Note that the Minkowskian correlation function is approached from
above in the complex plane of the hyperbolic angle $\chi$,
in agreement with the usual ``$-i\epsilon$''
prescription~\cite{Meggiolaro07}. 

According to the {\it crossing--symmetry relations}
(\ref{eq:crossrel}) (derived in~\cite{crossing}), we should find the
physical amplitude in the ``crossed'' channel 
at the edge  $E^{\rm
  cross}$ of the analyticity domain. Here we find 
\begin{equation}
  \overline{C}_{E\mu\nu}(\pi-i\chi, iT) =
  \sum_{\alpha,\beta=0}^3 iS_{\mu\alpha}S_{\nu\beta}{C}_{M\alpha\beta}(-\chi,T),
\end{equation}
where $S$ is a matrix which simply interchanges the $0$ and $1$
components of fields and coordinates, and which can be reabsorbed
into the loops with a further transformation of
fields and coordinates, with the only effect of reversing the
orientation of $W_{\Gamma_2}$, so that $W_{\Gamma_2}\to
W_{\Gamma_2}^*$. We thus find that the Euclidean correlation function 
is analytically continued to the physical correlation function (with 
positive hyperbolic angle $-\chi$) of a loop and an antiloop, as
expected~\cite{crossing}; as a by--product, we reobtain the
crossing--symmetry relation for the loops, Eq.~(\ref{eq:crossrel}),
\beq
{\mathcal{G}}_M(i\pi-\chi;T;\vec{z}_{\perp};1,2)
=\overline{\mathcal{G}}_M(\chi;T;\vec{z}_{\perp};1,\overline{2}) 
=\overline{\mathcal{G}}_M(\chi;T;\vec{z}_{\perp};\overline{1},2) ,
\quad \forall\chi\in {\cal I}_M, T\in\mathbb{R}^+.
\eeq{eq:cross_ext}
To see what happens at the other two edges of the analyticity domain
it suffices to recall that ${\cal D}_E$ possesses the
symmetry ${\cal D}_E={\cal D}_E^*$, and that the coefficients 
$\overline{C}_{E\mu\nu}$ satisfy the {\it reflection relation} 
 $ \overline{C}_{E\mu\nu}(z^*,T^*) = \overline{C}_{E\mu\nu}(z,T)^*$.
Exploiting this relation, $C$--invariance and reality of the
integration measure, one easily shows
that 
\begin{equation}
  \label{eq:reflG}
   \overline{\cal G}_E(z^*;T^*;\vec{z}_\perp;1,2) =  \overline{\cal
     G}_E(z;T;\vec{z}_\perp;1,2)^*.
\end{equation}
In particular, this means that at \mbox{$\psi=-\pi$} we find the
complex conjugate of the physical correlation functions, respectively
at  $E^{\rm dir}{}^*$ ($\chi<0$) for the ``direct channel'' and at
$E^{\rm cross}{}^*$ ($\chi>0$) for the ``crossed channel''. Moreover,
from the previous  relation we find that the Euclidean correlation
function at $\chi=0,\psi=0$ is a real function (see also ~\cite{lattice}).

\subsection{Analyticity properties of the correlation function with the
IR cutoff removed}

As the physically relevant quantities are the correlation functions
with the IR cutoff removed~\cite{BL,Meggiolaro05}, i.e., ${\cal
  C}_{M,E}$, 
we will discuss now what can be inferred about their analyticity
properties from the properties of ${\cal G}_E$.
As a function of the complex variable $T$ at fixed
$z= \theta-i\chi$, $\overline{\cal G}_E$ is analytic in the sector $-\pi/2 + \Delta < \arg T=\psi/2
< \Delta$, where $\Delta=\Delta(z)\in (0,\pi/2)$; 
one can then define
$I_\Delta\equiv (-\pi + 2\Delta,2\Delta)$, and rewrite ${\cal D}_E$ as
${\cal D}_E = \{(z,T)\,|\,z\in {\cal S}_E ,\,\psi\in I_{\Delta(z)}\}$.

Moreover, the normalised correlation function is expected to be IR
finite: in a non--Abelian gauge theory, the short--range nature of the
interactions implies that those parts of the partons' trajectories 
which lie too far aside with respect to the ``vacuum correlation
length''~\cite{DDSS} do not affect each other. There should then be a
``critical'' length $T_c$, beyond which the normalised correlation
function becomes independent of $T$: this is confirmed by lattice
calculations~\cite{lattice}. As the existence of a ``vacuum correlation
length'' is usually ascribed to the non--trivial dynamics dictated by
non--Abelian gauge invariance, the previous argument is expected to
apply also for the analytically--extended correlation functions,
substituting the real variable $T$ with the modulus of the complex
variable $|T|$. 

In conclusion, the analytically extended correlation
function is expected to satisfy the hypotheses of the
Phragm\'en--Lindel\"of theorem (see theorem 5.64 of Ref.~\cite{Tit}),
which implies that $\overline{\cal G}_E$ converges uniformly to a
unique value in the whole sector as $|T|\to\infty$. We can then define
unambiguously the function
\beq
  \overline{\cal C}_{E}(z;\vec{z}_\perp;1,2) \equiv \lim_{|T|\to\infty}
  \overline{\cal G}_{E}(z;T;\vec{z}_\perp;1,2),\qquad \forall z\in {\cal S}_{E}
\eeq{}
since the limit on the right--hand side does not depend on the
particular direction in 
which one performs it. One easily sees that 
$\overline{\cal C}_E$ is the analytic extensions of 
${\cal C}_E$, and 
if we now take the limit $|T|\to\infty$ in the
analytic continuation relation, Eq.~(\ref{eq:an_cont_phys}), we obtain the
analytic continuation relation with the IR cutoff removed~\cite{Meggiolaro05}, 
\beq
  {\cal C}_M(\chi;\vec{z}_\perp;1,2) =
  \overline{\cal C}_E(-i\chi;\vec{z}_\perp;1,2) ,
  \qquad \forall \chi\in {\cal I}_M.
\eeq{}
The crossing--symmetry relations are still valid for the analytic extensions
  $\overline{\cal C}_M$ and $\overline{\cal C}_E$ throughout the respective 
analyticity domains, as one can prove by
taking the limit  $|T|\to\infty$ in Eq.~(\ref{eq:cross_ext}) (relying
again on the Phragm\'en--Lindel\"of theorem mentioned above). Note
also that ${\cal C}_E(z^*)={\cal  C}_E(z)^*$ throughout the domain
of analyticity, as one can easily see by taking $|T|\to\infty$ in
Eq.~(\ref{eq:reflG}). 

\subsection{Lattice regularisation}

The functional integral must be regularised to
become a well--defined mathematical object; here we justify the
formal argument given above using a lattice regularisation. 
In this approach the ill--defined continuum functional integral is
replaced with a well--defined (multidimensional) integral, which in
the case of gauge theories can be chosen to be an integral on the
gauge group manifold~\cite{Wil}, 
\begin{equation}
  \label{eq:lfintegral}
  \la {\cal O}[U] \ra_{S_{\rm lat}} \equiv
  \frac{\displaystyle\int[DU]{\cal O}[U] e^{-S_{\rm
        lat}[U]}}{\displaystyle\int[DU]e^{-S_{\rm lat}[U]}}
\end{equation}
where $DU$ is the invariant Haar measure. 
It is easy to see that in our case the action
\begin{equation}
  \label{eq:mod_E_lat_ac}
  S_{\rm lat} = \beta\sum_{n,\mu < \nu}
  {C}_{E\mu\nu}(\theta,T) \left[1- \frac{1}{N_c}\Re\tr\,
    U_{\mu\nu}\right],
\end{equation}
where $U_{\mu\nu}$ is the usual plaquette variable (in the
fundamental representation)~\cite{Wil} and $\beta=2N_c/g^2$, gives
back the action $S_E^{\rm Y.M.}$ of Eq.~(\ref{eq:mod_E_ac}) in the
limit $a\to 0$, upon identification of the link variables with
$U_\mu(n)=\exp\{iga\phi_\mu(na)\}$. 
For compact gauge groups, such as $SU(N_c)$, the integration range is
compact, so that, as long as the volume and the lattice spacing are
finite, the integral (\ref{eq:lfintegral}) with the
action (\ref{eq:mod_E_lat_ac}) is convergent and analytic in $\theta$
and $T$; in  
the $V\to\infty$ limit, one has to impose positive--definiteness of
the real part of the action in order for the integral to remain
convergent, and this leads exactly to the {\it convergence conditions}
(\ref{eq:convcond}).

The action (\ref{eq:mod_E_lat_ac}) is correct at tree--level, but
one has also to ensure that quantum effects do not modify its
form. 
It is easy to see that Eq.~(\ref{eq:mod_E_lat_ac}) is also the
correct tree--level action for an anisotropic lattice regularisation
of the usual Euclidean Yang--Mills action, as one can directly check
[see Eqs.~(\ref{eq:ME-transf}) and (\ref{eq:matrix})] 
by identifying $U_\mu(n)=\exp\{iga_\mu A_{E\mu}(na)\}$, with
$a_\mu=a/M_{E\mu\mu}$. Showing that Eq.~(\ref{eq:mod_E_lat_ac}) is a
good lattice action on an isotropic lattice for the modified action
Eq.~(\ref{eq:mod_E_ac}) is then equivalent to show that it is a good
action on an anisotropic lattice for the usual Yang--Mills action.

Since the general anisotropic action 
is not guaranteed to belong to the same universality class as the
isotropic lattice action~\cite{Bur}, one has to enforce that rotation
invariance is restored in the continuum limit by 
properly tuning the coefficients of the various terms of the action,
obtaining in our case
\begin{equation}
  \label{eq:mod_E_lat_ac_2}
  \widetilde{S}_{\rm lat} = \sum_{n,\mu < \nu}
  \beta_{\mu\nu}{C}_{E\mu\nu}(\theta,T)
  \left[1- \frac{1}{N_c}\Re\tr \,U_{\mu\nu}\right],
\end{equation}
with properly chosen functions
$\beta_{\mu\nu}=\beta_{\mu\nu}(a,\theta,T)$. Due to the
asymptotic freedom property of non--Abelian gauge theories, one can
determine this functions analytically in perturbation
theory for small lattice spacings. We have calculated
$\beta_{\mu\nu}(a,\theta,T)$ to one--loop accuracy, and we have found
that quantum effects do not impose further restrictions on the
analyticity domain ${\cal D}_E$ found above~\cite{moi}. 

As a final remark, we notice that after the field and coordinate
transformation, the longitudinal sides of the two continuum 
Wilson loops are at $45^{\circ}$ with respect to the new axes, and
have to be approximated by a broken line (see, e.g., Ref.~\cite{lattice}): 
this introduces approximation errors which have to be carefully
considered, but which should vanish in the continuum limit, thus
leaving unaltered our analysis. One could use on--axis Wilson loops,
thus performing an ``exact'' calculation on the lattice, if one
performs a further transformation of the action, choosing the new
basis vectors along the directions of the longitudinal sides of the
loops, but in this case one has to deal with the more complicated
(``chair--like'') terms $\tr[U_{0\alpha_{\perp}}U_{1\alpha_{\perp}}^\dag]$.

\end{document}